\documentclass[10pt, conference, letterpaper]{IEEEtran}
\IEEEoverridecommandlockouts
\usepackage{import}
\usepackage{graphicx}
\usepackage{amsfonts}
\usepackage{amssymb}
\usepackage{amsmath}
\usepackage{tabularx}
\usepackage{multirow, makecell}
\usepackage[dvipsnames]{xcolor}
\usepackage{amsmath}
\graphicspath{ {./figures/} }
\usepackage{subcaption}
\usepackage{multirow}
\usepackage{textcomp}
\usepackage{algorithm}
\usepackage{algorithmic}

\ifCLASSOPTIONcompsoc
  \usepackage[nocompress]{cite}
\else
  \usepackage{cite}
\fi

\begin{document}

\title{5G New Radio Resource Allocation Optimization for Heterogeneous Services}


\author{\IEEEauthorblockN{Nasim Ferdosian}
\IEEEauthorblockA{ETIS UMR 8051\\
CY-Tech, ENSEA, CNRS\\
Cergy 95000 France\\
nasim.ferdosian@ensea.fr}
\and
\IEEEauthorblockN{Sara Berri}
\IEEEauthorblockA{ETIS UMR 8051\\
CY-Tech, ENSEA, CNRS\\
Cergy 95000 France\\
sara.berri@ensea.fr}
\and
\IEEEauthorblockN{Arsenia Chorti}
\IEEEauthorblockA{ETIS UMR 8051\\
CY-Tech, ENSEA, CNRS\\
Cergy 95000 France\\
arsenia.chorti@ensea.fr}
}

\maketitle
\begin{abstract}
5G new radio (NR) introduced flexible numerology to provide the necessary flexibility for multiplexing the communication of heterogeneous services on a shared channel. One of the fundamental challenges of 5G NR is to develop resource allocation schemes to efficiently exploit such flexibility to optimize resource allocation of ultra-reliable low-latency communications (URLLC) in coexistence with enhanced mobile broadband (eMBB) while ensuring their colliding performance requirements. To address this challenge, we present a new formulation of 5G NR resource allocation to accommodate eMBB and URLLC services, by considering their interplay. The objective of the formulated problem is to meet throughput quality of service (QoS) requirements probabilistically when the optimal global solution is infeasible. To this end, we express the problem as an integer linear program and consider two formulations with hard and soft URLLC throughput constraints. Furthermore, we propose a low-complexity two-step heuristic approach, inspired by an instance of bin packing optimization, to provide a trade-off between resource allocation efficiency and computational complexity. Finally, performance results are provided and demonstrate that the proposed approaches can provide low-complexity resource allocation solutions, in balance with the performance targets of the services.
\end{abstract}
\IEEEpeerreviewmaketitle

 \section{Introduction}\label{Intro}
 The emerging wide range of devices and services in a variety of industrial and societal fields have introduced new performance and quality of service (QoS) requirements to be addressed by 5G mobile communication systems \cite{ITU}. The main key enablers of 5G, as a service driven mobile communication technology is to address the requirements of the new era of heterogeneous services include flexible numerology, mini-slotting and optimized frame structure which have been defined by 5G New Radio (NR) \cite{3GPP}. Optimizing the allocation of 5G NR to support enhanced mobile broadband (eMBB), ultra-reliable and low latency communications (URLLC) services with varying QoS requirement and characteristics remains a challenging issue \cite{sadi2020flexible, pocovi2018multiplexing}. eMMB services require very high bit rates (Gigabits per second) and moderate latency (a few milliseconds) \cite{chen2018ultra}. URLCC services provide urgent and reliable data exchange with even submillisecond user plane latency. 

The current research body has introduced different preemption and puncturing approaches to multiplex both URLLC and eMBB services \cite{anand2020joint, pradhan2020joint, pedersen2017punctured, alsenwi2019embb}. In puncturing approaches, an arriving URLLC packet overtakes resources that have already been allocated to eMBB users, causes a throughput reduction to these kinds of data demanding services \cite{li2020deep}. Therefore, while the current research provides a variety of resource allocation approaches, there is currently a gap for an efficient resource allocation solution to meet throughput demands and latency constraints of URLLC services without compromising throughput of eMBB services. 

Alternatively, the authors in \cite{you2018resource} studied the resource allocation of eMBB and URLLC services, without using puncturing mechanisms to schedule resources, while exploiting the 5G NR flexible numerology and frame structure. Exploiting this flexibility to optimize the resource allocation to different services while ensuring their QoS requirements, was shown to be an $NP$-hard problem. They proposed different optimization methods to solve the problem however, a trade-off between resource allocation efficiency and computational complexity is still a challenging issue. 


In this article, we study the problem of 5G NR resource allocation to URLLC services coexisted with eMMB services. We formulate this problem as an integer linear programming problem, where the objectives is to determine resource allocation decisions that maximize the sum throughput of the eMBB services, subject to the throughput demands and latency constraints of the URLLC services. Moreover, we propose an alternative formulation of the problem with less stringent constraints to improve the feasibility of the problem.
The overall problem is then treated as a specific instance of bin packing optimization, whose objective is to minimize the placements of URLLC services. We propose a greedy scheduling algorithm as an approximate solution to the problem. Finally, we conduct a numerical analysis, and show that the proposed heuristic scheduling algorithm provides a near-optimal solution in polynomial-time. In addition, we evaluate the optimal solutions of the two different formulations, and show that gains could be achieved by avoiding a close relationship between the eMBB and URLLC services.

The rest of the paper is organized as follows. We present an overview of the flexible numerology and frame structure in 5G NR in Section \ref{Sys}. In Section \ref{Prob} we give the problem, while in Section \ref{Alg} we introduce a re-formulation of the problem and a light-weight near-optimal heuristic. In Section \ref{Resu} we present an extensive set of numerical results and related analysis. Finally, Section \ref{Conc} concludes the paper.


\section{Flexible Numerology in 5G NR}\label{Sys}
5G NR Release-15 \cite{3GPP} defines a flexible numerology with subcarrier spacing (SCS) of $15$, $30$, and $60$ kHz below $6$ GHz, and $60$ and $120$ kHz above $6$ GHz, compared to long-term evolution (LTE) which uses a fixed numerology with SCS of $15$ kHz below $6$ GHz. 5G NR also defines a $10$ milliseconds (ms) frame, with each frame divided into $10$ subframes of $1$ ms, which are further divided into one or more mini-slots. A mini-slot comprises $14$ OFDM symbols for a configuration using normal cyclic prefix, or $12$ OFDM symbols for extended cyclic prefix. In 5G NR, the mini-slot size is defined according to the symbol duration, which is inverse to the SCS, to ensure the orthogonality of the subcarriers. By using higher SCS, the symbol duration decreases and hence also the mini-slot size, which is beneficial for lower latency \cite{semiari2019integrated}.

\begin{table}[!htbp]
 \renewcommand{\arraystretch}{1.2}\centering \caption{Resource Blocks in Flexible Numerology}
\setlength{\tabcolsep}{6pt}
\centering \label{ta:flexible numerology}
\begin{tabular}{|l|c|c|c|c|}
\hline
\multicolumn{1}{|c|}{} &  \multicolumn{1}{c|}{Shape 1} & \multicolumn{1}{c|}{Shape 2} & \multicolumn{1}{c|}{Shape 3} & \multicolumn{1}{c|}{Shape 4}\\
\hline
TTI duration (ms) & 0.5 & 0.25 & 0.125 & 0.125\\
\hline
SCS (kHz) & 15 & 30 & 60 & 60\\
\hline
Symbol duration ($\mu$s) & 66.7 & 33.3 & 16.7 & 16.7\\
\hline
CP ($\mu$s) & 4.7 & 2.3 & 1.2 & 4.17\\
\hline
Number of Symbols & 7 & 7 & 7 & 6\\
\hline
\end{tabular}
\end{table}  


In the present paper, we focus on a downlink resource scheduling scenario, where one base station (BS) serves both throughput hungry (eMMB) and ultra-low latency (URLLC) users \cite{you2018resource}. A resource allocation to a user consists in assigning to it a set of adjacent SCSs and mini-slots in the time-frequency grid, referred to as resource blocks. The permissible shapes of the resource blocks depend on the numerology and frame structure employed and can be either fixed or dynamically chosen, this latter case being dubbed as flexible numerology. Table \ref{ta:flexible numerology} presents the four most widely utilized resource block shapes determined by different configurations of numerology and frame structure, according to the 5G NR specifications. According to the flexible numerology principle, no restrictions are placed to eMBB and URLLC users with respect to the shape of resource blocks utilized to serve them.

In this framework, to optimize eMBB and URLLC coexistence, a joint scheduler allocates corresponding resource blocks in the time-frequency grid with the objective to maximize the sum throughput of the former, while satisfying the throughput demands and latency constraints of the latter. Current proposed solutions using puncturing or preemptive scheduling have been shown to incur heavy losses for eMBB users. Furthermore, previous works, identifying on the fly the optimal solution of the corresponding combinatorial optimization problem, have shown that depending on the size of the scheduling grid, the affordable latency and the throughput demands of URLLC users, a solution might not always be feasible. 

To address such shortcomings, in the present work we propose a reformulation of the standard scheduling problem, referred to as P$1$ in the following, that allows  ``dropping'' some URLLC services in such cases of infeasibility, i.e., we propose to allow for partial coverage of URRLC demands whenever unable to respect the whole set of constraints. This work moves towards the direction of assigning  a differentiated services (DiffServ) typeof \textit{throughput QoS guarantees}  to layer $2$ radio access network (RAN) scheduling, a task typically performed at layer 3. The motivation behind this undertaking is that as network slicing and network function virtualization will be handled by the virtualization orchestrator, flexible QoS  guarantees will be jointly handled by a unique virtualization layer.

Next, we first present the original scheduling problem with both hard demand and hard delay constraints for URLLC users, followed by a  re-formulation with soft demand and delay constraints for URLLC users. Finally, we discuss a novel heuristic scheduler that solves the original P$0$ problem on a best effort approach for each individual URLLC user, i.e., near-optimal scheduling is proposed whenever possible, otherwise (some of) the URLLC demands are (partially) dropped.

\section{Scheduling Problem Formulation with Hard Throughput Constraints}\label{Prob}
In the following, let us denote by $\mathcal{K}$ the set of all users, by $\mathcal{K}^{(c)}$ the set of eMBB users and by $ \mathcal{K}^{(\ell)} $ the set of URLLC users. $\mathcal{B} $ denotes the set of all possible resource blocks according to the employed numerology and finally,  $ \mathcal{I} $ denotes the set of all mini-slots. We utilize the  indicator $\alpha_{b,i}, b\in \mathcal{B}, i \in \mathcal{I}$ which is set to unity when a block $b\in \mathcal{B}$ includes basic unit $i\in \mathcal{I}$, in which case $\alpha_{b,i}=1$, otherwise $\alpha_{b,i}=0$. Furthermore, we denote by $r_{b,k}, b\in \mathcal{B}, k \in \mathcal{K}$ the achievable throughput of user $k$ over the resource block $b$. 
Under the URLLC latency constraint, the resource block $b$ cannot be assigned to $ k\in \mathcal{K}^{(\ell)} $ if the end time of block $b$ exceeds its latency constraint; this is embedded as a hard constraint by imposing $r_{b,k}=0$ for respective resource blocks $b \in \mathcal{B}$ and user $k \in \mathcal{K}^{(l)}$. Finally, by $x_{b,k}$ we denote a binary variable that takes the value $1$ if the resource block $b\in \mathcal{B}$ is assigned to user $k\in \mathcal{K}$, otherwise $x_{b,k}=0$.
 
The standard scheduling optimization problem is to maximize the sum throughput of $\mathcal{K}^{(c)}$ users under the constraint of satisfying the latency and throughput demands of the set  $\mathcal{K}^{(\ell)}$, without any overlapping between the allocated blocks. In other words, our goal is to find the resource allocation that satisfies the URLLC users' demands, with minimal losses for eMBB users in terms of throughput, and, subsequently schedule all the remaining resource blocks to the eMBB users. The formal problem formulation is given as \cite{you2018resource}
\begin{align}
\text{[P0]} \quad \max_{x_{b,k} \in \{0,1\}} &  \sum_{b\in\mathcal{B}}\sum_{k\in\mathcal{K}^{(c)}} r_{b,k}x_{b,k}, \label{eq:objective}\\
\text{s.t.} \quad & \sum_{b\in\mathcal{B}}r_{b,k}x_{b,k}\geq q_{k}, \quad k\in\mathcal{K}^{(\ell)} \label{eq:constraint1},\\
           & \sum_{b\in\mathcal{B}}\sum_{k\in\mathcal{K}}\alpha_{b,i}x_{b,k}\leq 1, \quad i\in\mathcal{I} \label{eq:constraint2}.
\end{align}

In \cite{you2018resource} it was proven that the combinatorial problem P0 is an ${NP}$-hard partition problem. Furthermore, the superior performance of scheduling using flexible as opposed to fixed numerology was demonstrated and established. Albeit, though an extensive set of results the authors also showed that for certain sets of parameters, P$0$ becomes infeasible; in particular, as the delay constraints of URLLC users become more stringent and / or their throughput demands increase, it might be impossible to find a solution in fixed size time-frequency grids.


\begin{figure}[!htbp]
    \centering
    \includegraphics[width=1.00\columnwidth]{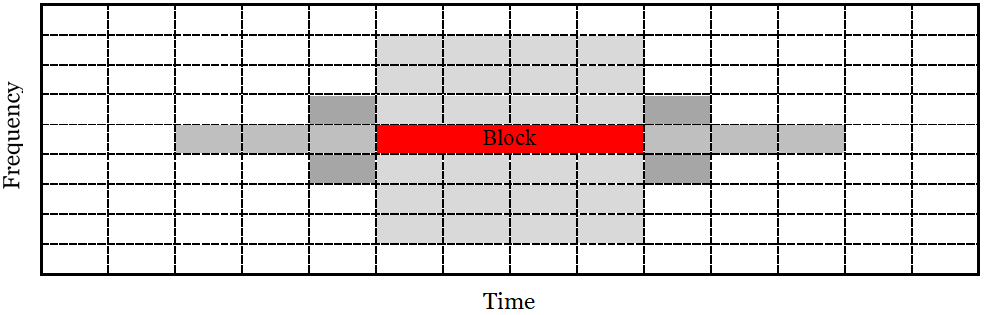}    
    \caption{Resource allocation of a candidate block and its corresponding conflicts; vertical blocks (light grey), horizontal blocks (grey) and square blocks (dark grey).}
\end{figure}

\par On a different note, the authors in \cite{ourINFOCOM} investigated an alternative formulation of P$0$ by introducing an explicit description of the impact constraint (\ref{eq:constraint2}), i.e., of the fact that the resource blocks are not allowed to overlap, to the optimal solution. To this end they introduced the concept of ``conflict" that incurs by any specific URLLC or eMBB resource block placement to subsequent placements. To illustrate the idea, Fig. 1 depicts all the ``conflicts" that arise from an arbitrary block placement, shown in red; the specific block allocation (in red) \textit{forbids} any other block allocation in the sketched neighborhood (in grey), as constraint (\ref{eq:constraint2}) does not allow for overlapping (partial or full) of block placements. In light of this, it is evident that even if a particular resource block might achieve maximum throughput, its allocation can be sub-optimal due to the losses because of the generated forbidden placements around it, i.e., the generated conflicts described by constraint (\ref{eq:constraint2}) of P$0$ might outweigh the throughput gains. Leveraging this insight, a bin packing type of formulation of the P$0$ was proposed and related near-optimal heuristics were proposed.

In the following, we will first provide a reformulation of P$0$ that allows for probabilistic coverage of the URLLC demands in view of the fact that P$0$ is infeasible in certain scenarios, dubbed as P$1$. Subsequently, inspired by both P$1$ and the bin packing formulation  proposed in \cite{ourINFOCOM}, a novel heuristic scheduler is presented to accommodate URLLC user demands on a best effort approach.

\section{Problem Reformulation and Heuristic Scheduler}

\subsection{Reformulation of P$0$ with soft URLLC throughput constraints}
The solution of the problem P0 mainly depends on the constraints \eqref{eq:constraint1}, which require that all the URLLC users' demands should be satisfied. Nevertheless, despite the desirable benefits brought by this formulation to the URLLC users, the constraints invoked raise the big issue of infeasibility when the demand is not fully covered, leading to not scheduling at all the eMBB users as well\footnote{Following the strict formulation of P$0$, when the URLLC placements are infeasible, the scheduler does not allocate any resources at all.}. Indeed, it is important, and even necessary, to address the shortcoming of this formulation, by introducing the possibility of dropping (some of) the URLLC services whenever necessary. 

To this end, we propose in the following an alternative formulation that aims at maximizing the joint overall sum throughput, i.e., including both URLLC and eMBB services, with less stringent constraints for satisfying the throughput demands of URLLC users. The problem re-formulation, dubbed as P$1$ is given below:


\begin{align}
\text{[P1]} \quad \max_{x_{b,k} \in \{0,1\}} &  \sum_{b\in\mathcal{B}}\sum_{k\in\mathcal{K}^{(c)}} r_{b,k}x_{b,k}, \label{eqOF}\\
\text{s.t.} \quad & \sum_{b\in\mathcal{B}}r_{b,k}x_{b,k}\leq q'_{k}, \quad k\in\mathcal{K}^{(\ell)} \label{eqC1},\\
          & \sum_{b\in\mathcal{B}}\sum_{k\in\mathcal{K}}\alpha_{b,i}x_{b,k}\leq 1, \quad i\in\mathcal{I} \label{eqC2}.
\end{align}

where: $q'_k=q_k+u_k, k \in \mathcal{K}^{(l)}$ and $u_k\geq0$. The parameter $u_k$ is introduced to assign more or less importance to the URLLC services. Moreover, it might depend on the throughput (and by extension to the delay) tolerance. 

The key idea of the proposed re-formulation is to take advantage of the fact that the feasibility of the two services and the different demands could be treated independently; this would make it possible to perform  placement for the eMBB services without having as a prerequisite to satisfy all of the constraints of the URLLC users. Note that the proposed re-formulation is compromising the URLLC users' demands as these are also considered in the objective to avoid the trivial solution $x^{\star}_{b,k}=0, k \in \mathcal{K}^{(l)}$, $\forall b \in \mathcal{B}$, which is not beneficial.

\subsection{Heuristic Scheduler}\label{Alg}
In \cite{ourINFOCOM}, it has been demonstrated that to maximize the throughput of $\mathcal{K}^{(c)}$ services while accommodating the data rate demands and latency constraints of $\mathcal{K}^{(\ell)}$ users, it is necessary, on average, to minimize the number of conflicts. In turn, the minimization of the number of conflicts can be  achieved in principle through the minimization of the number of URLLC placements, which points to casting the scheduling problem as a bin packing optimization problem. 

Motivated by this result, here, we propose a novel, computationally efficient scheduling approach, inspired by the Harmonic-k heuristic \cite{schreiber2013improved} 
for the standard bin packing problem. 
The proposed resource allocation heuristic solution, summarized in Algorithm \ref{alg1}, performs allocation of blocks to $\mathcal{K}^{(\ell)}$ services based on the minimization of the number of conflicts which is achieved by the minimizing the number of $\mathcal{K}^{(\ell)}$ resource allocations (placements). To this end, Algorithm \ref{alg1} finds an allocation that satisfies the $\mathcal{K}^{(\ell)}$s' demands by using the least number of resource blocks. 
The resource allocation for $\mathcal{K}^{(\ell)}$ services and $\mathcal{K}^{(c)}$ services are treated in two separate phases. The former is performed first because of the latency requirement.



\begin{algorithm}[t]
\caption{Bin Packing Resource Allocation Algorithm}
\label{alg1}
\begin{algorithmic}
\REQUIRE throughput matrix $\mathbf{r}=[r_{b,k}], \quad b\in \mathcal{B}, k\in \mathcal{K}$, aggregated-throughput-loss vector $\mathbf{e}$, demand vector of URLLC services $\mathbf{q}$, set of all available resource blocks $\mathcal{B}$.
\ENSURE Block-service assignment $\mathbf{s}$.

\FOR{$ k=1$ to $|\mathbf{q}|$} 
\STATE create the following categories:
\FOR{$ i=1$ to $M$} 
\STATE $Cat^i U^k=$ all resource blocks $b \in \mathcal{B}$ where \\ $\lceil{q_{k}/r_{b,k}}\rceil = i$;
\STATE Check pairwise conflicts among categorized blocks and remove the blocks with the higher aggregated-throughput-loss;
\ENDFOR
\ENDFOR
\\
\textbf{Phase ($\mathcal{K}^{(\ell)}$ resource allocation):}
\FOR{$i=1$ to $M$} 
\STATE select the $Cat^i U^k$ which has the least number of blocks;
\IF{ ($|Cat^i U^k| \geq i$ and $q_{k}$ is not already met)}
\STATE $ \mathcal{B} ^ \prime \leftarrow $ (select $i$ number of blocks in $Cat^i U^k$ with the least aggregated-loss-value);
\STATE	$\mathbf{s} \leftarrow \mathbf{s} \cup {(b ^ \prime, k ^ \prime)}$ , $k ^ \prime=i , \forall b ^ \prime \in \mathcal{B} ^ \prime$;
\STATE Remove from $\mathcal{B}$ the blocks in $\mathbf{s}$ and those overlapping with the blocks in $\mathbf{s}$;

\IF{ $q_{k}$ is met}
\STATE $\mathcal{K}^{(\ell)} \leftarrow \mathcal{K}^{(\ell)} \backslash \{k ^ \prime \}$;
\ENDIF
\ENDIF
\ENDFOR
\\
\textbf{Phase ($\mathcal{K}^{(c)}$ resource allocation):}
\REPEAT 
\STATE $(b ^ \prime, k ^ \prime) \leftarrow \arg\max_{ b \in \mathcal{B}, k \in \mathcal{K}^{(c)}}  r_{b,k}$;
\STATE  $\mathbf{s} \leftarrow \mathbf{s} \cup {(b ^ \prime, k ^ \prime)}$;
\STATE Remove from $\mathcal{B}$ the blocks in $\mathbf{s}$ and those overlapping with the blocks in $\mathbf{s}$;
\UNTIL{ $\mathcal{B} = \emptyset$ }

\end{algorithmic}
\end{algorithm}

 
 For each $k \in \mathcal{K^{(\ell)}}$ we generate $M$ categories with decreasing fractional sizes with respect to $q_k, k \in \mathcal{K^{(\ell)}}$, i.e., category $i\in \{1,\ldots,M\}$ is defined as the set of all
resource blocks $b \in \mathcal{B}$ for which the ceiling of the ratio of the service demand over the throughput of block $b$ is equal to $i$, or equivalently, category $Cat^i U^k$ contains the available resource blocks which satisfy at least $1/i$-th of the service demand $q_k$. 

 For example, $Cat^1 U^1$ is the category of the blocks which individually satisfy the whole service $1$'s demand. 
Then, in $\mathcal{K}^{(\ell)}$ resource allocation phase, the minimization of the number of $\mathcal{K}^{(\ell)}$ placements is achieved by starting from the categories of largest blocks, i.e., blocks in $Cat^1 U^k$ are allocated first and then we move to $Cat^2U^k$, and so on. 
The elements of each selected category $Cat^i U^k$, are then ordered with decreasing aggregate loss for the eMBB users, on operation of complexity $|Cat^iU^k| \log |Cat^iU^k| $ and the first $i$ of them are selected for the placement of URLLC user $k$. If there are not enough elements for the allocation, then the respective blocks are moved the next in chain category. 

After each placement the allocated blocks are removed from $\mathcal{B}$ and all other categories. In the case that a URLLC user cannot be accommodated, then no blocks are allocated to it. 

Once the set of URLLC users has been serviced, we move to the last phase of the algorithm for the resource allocation to $\mathcal{K}^{(c)}$ services. Algorithm 1 selects the block-service pairs with the highest throughput $r_{b,k}, k \in \mathcal{K}^{(c)}$ from the remaining available resource blocks and continues until no more blocks are available. 

\section{Numerical Results}\label{Resu}
\subsection{Performance Comparison of P$0$, P$1$ and of the Heuristic Scheduler in Terms of eMBB Sum Throughput}

To showcase the effectiveness of the proposed re-formulation of the resource allocation as P$1$ and of Algorithm \ref{alg1}, we present numerical results for the sum throughput of the eMBB services 
using the same simulation setup as in \cite{you2018resource}\footnote{We thank the authors of \cite{you2018resource} for kindly sharing their simulation codes in IEEE DataPort.}. This simulation environment was implemented based on  the control channel overhead model for supporting the flexible numerology defined in \cite{miao2017physical} and considers the effect of guard band (i.e., of the cyclic prefix) on the achievable data rate by resource blocks as modeled in \cite{yazar2018flexibility}.
The global optimum solution of P$0$ is calculated by employing Gurobi optimization solvers\footnote{ Solving optimally P$0$ is non-scalable  with high complexity  and is used for benchmarking}. In addition, the global optimum solution of P$1$ is obtained by using the optimization solver IBM ILOG CPLEX. 

The parameter $u_k$ is varying with respect to the latency tolerance as follows: in case of URLLC bit rate demands 64 kbps and 128 kbps it set to 136, 116 and 96 for latency $\in\{0.25, 1\}$, $\in\{0.5\}$, and  $\in\{1.5, 2\}$, respectively; in case of URLLC bit rate demands 256 kbps it set to 244, and 124 for latency $\in\{0.25, 0.5, 1\}$, and  $\in\{1.5, 2\}$, respectively; in case of URLLC bit rate demands 512 kbps it set to 158, and 138 for latency $\in\{0.25, 0.5, 1, 1.5\}$, and  $\in\{2\}$, respectively; in case of URLLC bit rate demands 1024 kbps it set to 176 for latency $\in\{0.25, 0.5, 1, 1.5,2\}$. 

Different 5G numerologies and URLLC configurations in terms of data rate demands and latency tolerances are considered. In more detail, we 
considered URLLC latency constraints $\mathbf{\tau}=\{0.25, 0.5, 1, 1.5, 2\}$ msec and throughput demands ${q_k}=\{16, 32, 64, 128, 256, 512, 1024\}$ kbps for a set of $|\mathcal{K}^{(\ell)}|=5$ 
services $k \in \mathcal{K}^{(\ell)}$. The obtained results are presented in Figs. 2-6. The bit rates per user in $k \in \mathcal{K}^{(c)}$ for all the examined algorithms in case of URLLC bit rate demands $16$ kbps and $32$ kbps are almost same as the ones in case of data demand $64$ kbps. Therefore, for the sake of brevity, we omit the presentation of this set of results.


Beginning the discussion with the performance of P$1$, the numerical results show that 
as the latency tolerance decreases and the throughput demands of $\mathcal{K}^{(\ell)}$ users increase, in contrast to P$0$ which becomes infeasible, the formulation P1 always provides a solution irrespective of the parameters. This is thanks to the fact that P1 aims at a global sum throughput optimization with relaxed constraints for the URLCC throughput coverage. However, in some cases (e.g $\tau=0.5$ ms and bit rate 256 kbps), P1 fails to satisfy the URLLC demands whereas this is feasible using P0. In these cases however, a higher sum throughput for $K^{(c)}$ services is proposed. Interestingly, when the throughput demand is small (e.g., 64 kbps), all the URLLC demands are satisfied and the gap between the throughput of the set of $\mathcal{K}^{(c)}$ users provided by the two formulations, P0 and P1, is relatively small when the latency is either $\tau=1.5$ ms or $2$ ms and $0$ otherwise. These results suggest that, it is possible to consider either P1 or P0 without performance loss even when P0 is feasible. Furthermore, the formulation P1 could provide a better trade-off.

\begin{figure}[!htbp]
\centering
\vspace{-0.0em}
\includegraphics[width=1.00\columnwidth]{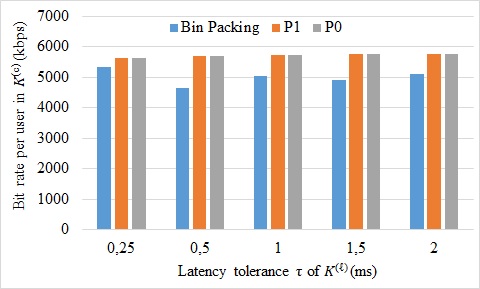}
\vspace{-0.8em}
\caption{Bit rate of $\mathcal{K}^{(c)}$ with respect to latency tolerance of $\mathcal{K}^{(\ell)}$ for the bit rate demand of $\mathcal{K}^{(\ell)}$ equals 64 kbps.} 
\label{fig:64kbps}
\vspace{-0.0em}
\end{figure}

\begin{figure}[!htbp]
\centering
\vspace{-0.0em}
\includegraphics[width=1.0\columnwidth]{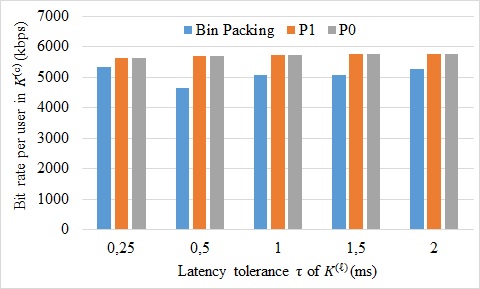}
\vspace{-0.8em}
\caption{Bit rate of $\mathcal{K}^{(c)}$ with respect to latency tolerance of $\mathcal{K}^{(\ell)}$ for the bit rate demand of $\mathcal{K}^{(\ell)}$ equals 128 kbps.} 
\label{fig:128kbps}
\vspace{-0.0em}
\end{figure}

\begin{figure}[!htbp]
\centering
\vspace{-0.0em}
\includegraphics[width=1.00\columnwidth]{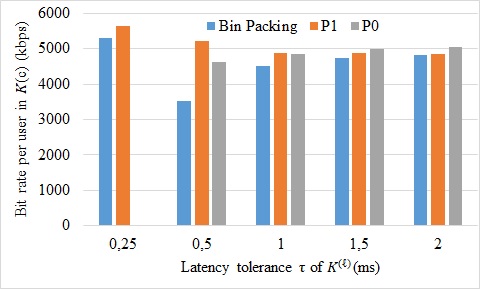}
\vspace{-0.8em}
\caption{Bit rate of $\mathcal{K}^{(c)}$ with respect to latency tolerance of $\mathcal{K}^{(\ell)}$ for the bit rate demand of $\mathcal{K}^{(\ell)}$ equals 256 kbps.} 
\label{fig:256kbps}
\vspace{-0.0em}
\end{figure}

\begin{figure}[!htbp]
\centering
\vspace{-0.0em}
\includegraphics[width=1.00\columnwidth]{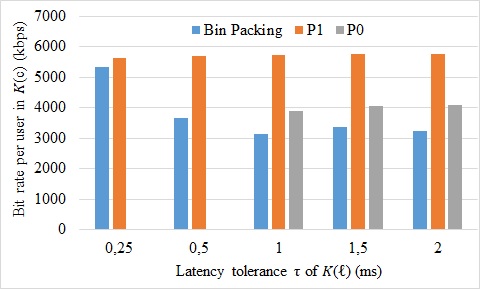}
\vspace{-0.8em}
\caption{Bit rate of $\mathcal{K}^{(c)}$ with respect to latency tolerance of $\mathcal{K}^{(\ell)}$ for the bit rate demand of $\mathcal{K}^{(\ell)}$ equals 512 kbps.} 
\label{fig:512kbps}
\vspace{-0.0em}
\end{figure}

\begin{figure}[!htbp]
\centering
\vspace{-0.0em}
\includegraphics[width=1.00\columnwidth]{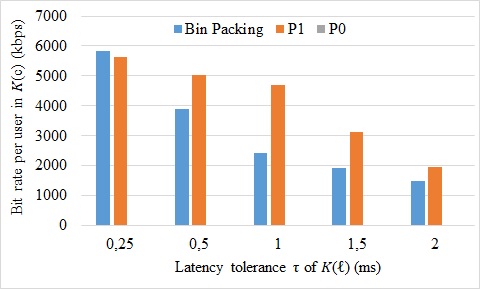}
\vspace{-0.8em}
\caption{Bit rate of $\mathcal{K}^{c}$ with respect to latency tolerance of $\mathcal{K}^{(\ell)}$ for the bit rate demand of $\mathcal{K}^{(\ell)}$ equals 1024 kbps.} 
\label{fig:1024kbps}
\vspace{-0.0em}
\end{figure}

With respect to the proposed heuristic scheduler, as can be seen in Figs. \ref{fig:64kbps}-\ref{fig:1024kbps}, it incurs negligible performance loss when P0 is feasible and succeeds in providing an allocation in all scenarios. Comparing the heuristic scheduler to P1, the sum throughput of eMBB users is lower in most of the cases. However, as will be discussed next, it manages to cover the full URLLC demands for wider palette of delay and throughput parameters for the URLLC users.



\subsection{Performance  Comparison  of  P0,  P1 and  of  the  Heuristic Scheduler in Terms of URLLC Coverage}\label{Resu0}
In this subsection, we compare the results of P0, P1 and the heuristic scheduler in terms of URLLC coverage. For this, we show in Table \ref{ratio}, the percentage of covered URLLC services for all the considered algorithms. From the table, we can see that the heuristic scheduler covers all the URLLC almost all the time, and always provides better results than the formulations P0 and P1. This is because, the heuristic schedules the two types of services sequentially, and the URLLC services are performed first, which explains consequently its lowest performance in terms of sum throughput of the eMBB services, when the percentage of covered URLLC is not zero (all the considered cases except for: $\tau=0.25$ and bit rate 1024).  On the other hand, when the coverage of all the URLLC services is feasible the formulation P0 is more favorable than P1 because it includes hard URLLC throughput constraints, which avoid to drop URLLC services. 
\begin{table}[h!]
\caption{Satisfied URLLC demands ratio in \%.}
\begin{center}
\begin{tabular}{|c|c|c|c|c|c|}
\hline
 $\tau$&Problem&64 kbps&256 kbps&512 kbps&1024 kbps  \\ \hline \hline
\multirow{3}{*}{0.25 ms}
&Bin Packing&100\%&80\%&40\%&0\% \\ \cline{2-6}
&P1&100\%&0\%&0\%&0\% \\ \cline{2-6}
&P0&100\%&0\%&0\%&0\%\\ \cline{1-6}

\multirow{3}{*}{0.5 ms}
&Bin Packing&100\%&100\%&100\%&40\% \\ \cline{2-6}
&P1&100\%&40\%&20\%&0\% \\ \cline{2-6} 
&P0&100100\%&100\%&0\%&0\% \\ \cline{1-6}

\multirow{3}{*}{1 ms}
&Bin Packing&100\%&100\%&100\%&100\% \\ \cline{2-6}
&P1&100\%&80\%&40\%&0\% \\ \cline{2-6} 
&P0&100\%&100\%&100\%&0\% \\ \cline{1-6}

\multirow{3}{*}{1.5 ms}
&Bin Packing&100\%&100\%&100\%&100\% \\ \cline{2-6}
&P1&100\%&100\%&100\%&20\%\\ \cline{2-6}
&P0&100\%&100\%&100\%&0\% \\ \cline{1-6}

\multirow{3}{*}{2 ms}
&Bin Packing&100\%&100\%&100\%&100\% \\ \cline{2-6}
&P1&100\%&100\%&100\%&60\% \\ \cline{2-6}
&P0&100\%&100\%&100\%&0\% \\ \cline{1-6}
\end{tabular}
\label{ratio}
\end{center}
\end{table}

\section{Conclusion}\label{Conc}
In this paper, we have studied the joint URLLC and eMBB resource allocation, over the envisioned flexible physical layer architecture for 5G, by accounting for the conflicts. Our goal is to determine the resource allocation to meet throughput demands and latency constraints of URLLC services without compromising eMBB throughput. We formulated the global problem as an integer linear program with hard URLLC throughput constraints, and then proposed to re-formulate the considered problem with less stringent constraints to improve the feasibility of the problem. Moreover, a heuristic solution is proposed which solves the problem by disassociating the URLLC and eMBB services and scheduled the former first to meet their latency requirements. Numerical results have demonstrated that, \itshape i)~\upshape the new re-formulation with soft URLLC throughput constraints is always feasible which is not the case when we consider hard constraints, and,
\itshape ii)~\upshape our proposed heuristic is near-optimal even better when we consider hard URLLC throughput constraints. \itshape ii)~\upshape in terms of URLLC coverage, the proposed heuristic performs better than the optimal solutions which cannot always produce feasible resource allocation. In future work, other heuristic solutions that meet probabilistic QoS requirements of other kinds of services, such as massive machine-type communications, at high load conditions of network traffic, will be investigated.



\bibliographystyle{IEEEtran}
\bibliography{biblio.bib}

\begin{thebibliography}{10}
\providecommand{\url}[1]{#1}
\csname url@samestyle\endcsname
\providecommand{\newblock}{\relax}
\providecommand{\bibinfo}[2]{#2}
\providecommand{\BIBentrySTDinterwordspacing}{\spaceskip=0pt\relax}
\providecommand{\BIBentryALTinterwordstretchfactor}{4}
\providecommand{\BIBentryALTinterwordspacing}{\spaceskip=\fontdimen2\font plus
\BIBentryALTinterwordstretchfactor\fontdimen3\font minus
  \fontdimen4\font\relax}
\providecommand{\BIBforeignlanguage}[2]{{%
\expandafter\ifx\csname l@#1\endcsname\relax
\typeout{** WARNING: IEEEtran.bst: No hyphenation pattern has been}%
\typeout{** loaded for the language `#1'. Using the pattern for}%
\typeout{** the default language instead.}%
\else
\language=\csname l@#1\endcsname
\fi
#2}}
\providecommand{\BIBdecl}{\relax}
\BIBdecl

\bibitem{ITU}
ITU, ``{5G Overview},'' {Setting the Scene for 5G: Opportunities and
  Challenges}, Geneva: International Telecommunication Union (ITU), 2018.

\bibitem{3GPP}
3GPP, ``{NR; Physical channels and modulation},'' {3rd Generation Partnership
  Project (3GPP)}, Technical Specification (TS) 38.211, 03 2020, version
  16.1.0.

\bibitem{sadi2020flexible}
Y.~Sadi, S.~Erkucuk, and E.~Panayirci, ``{Flexible Physical Layer Based
  Resource Allocation for Machine Type Communications Towards 6G},'' in
  \emph{2020 2nd 6G Wireless Summit (6G SUMMIT)}.\hskip 1em plus 0.5em minus
  0.4em\relax IEEE, 2020, pp. 1--5.

\bibitem{pocovi2018multiplexing}
{Pocovi, Guillermo and Pedersen, Klaus I and Mogensen, Preben}, ``Multiplexing
  of latency-critical communication and mobile broadband on a shared channel,''
  in \emph{2018 IEEE Wireless Communications and Networking Conference
  (WCNC)}.\hskip 1em plus 0.5em minus 0.4em\relax IEEE, 2018, pp. 1--6.

\bibitem{chen2018ultra}
H.~Chen, R.~Abbas, P.~Cheng, M.~Shirvanimoghaddam, W.~Hardjawana, W.~Bao,
  Y.~Li, and B.~Vucetic, ``{Ultra-Reliable Low Latency Cellular Networks: Use
  Cases, Challenges and Approaches},'' \emph{IEEE Communications Magazine},
  vol.~56, no.~12, pp. 119--125, 2018.

\bibitem{anand2020joint}
A.~Anand, G.~De~Veciana, and S.~Shakkottai, ``{Joint Scheduling of URLLC and
  eMBB Traffic in 5G Wireless Networks},'' \emph{IEEE/ACM Transactions on
  Networking}, vol.~28, no.~2, pp. 477--490, 2020.

\bibitem{pradhan2020joint}
A.~Pradhan and S.~Das, ``{Joint Preference Metric for Efficient Resource
  Allocation in Co-Existence of eMBB and URLLC},'' in \emph{Proc. Int. Conf.
  Commun. Syst. \& Netw. (COMSNETS)}.\hskip 1em plus 0.5em minus 0.4em\relax
  IEEE, 2020, pp. 897--899.

\bibitem{pedersen2017punctured}
K.~I. Pedersen, G.~Pocovi, J.~Steiner, and S.~R. Khosravirad, ``{Punctured
  Scheduling for Critical Low Latency Data on a Shared Channel with Mobile
  Broadband},'' in \emph{Proc. 86th Veh. Technol. Conf. (VTC-Fall)}.\hskip 1em
  plus 0.5em minus 0.4em\relax IEEE, 2017, pp. 1--6.

\bibitem{alsenwi2019embb}
M.~Alsenwi, N.~H. Tran, M.~Bennis, A.~K. Bairagi, and C.~S. Hong, ``{eMBB-URLLC
  Resource Slicing: A Risk-Sensitive Approach},'' \emph{IEEE Commun. Lett.},
  vol.~23, no.~4, pp. 740--743, 2019.

\bibitem{li2020deep}
J.~Li and X.~Zhang, ``{Deep Reinforcement Learning Based Joint Scheduling of
  eMBB and URLLC in 5G Networks},'' \emph{IEEE Wireless Communications
  Letters}, 2020.

\bibitem{you2018resource}
L.~You, Q.~Liao, N.~Pappas, and D.~Yuan, ``{Resource Optimization with Flexible
  Numerology and Frame Structure for Heterogeneous Services},'' \emph{IEEE
  Communications Letters}, vol.~22, no.~12, pp. 2579--2582, 2018.

\bibitem{semiari2019integrated}
O.~Semiari, W.~Saad, M.~Bennis, and M.~Debbah, ``{Integrated Millimeter Wave
  and Sub-6 GHz Wireless Networks: A Roadmap for Joint Mobile Broadband and
  Ultra-Reliable Low-Latency Communications},'' \emph{IEEE Wireless
  Communications}, vol.~26, no.~2, pp. 109--115, 2019.

\bibitem{ourINFOCOM}
N.~Ferdosian, S.~Skaperas, A.~Chorti, and L.~Mamatas, ``{Near Optimal Linear
  Complexity Scheduling of Heterogeneous Services by Resolving Conflicts},'' in
  \emph{Submitted in IEEE INFOCOM 2021-IEEE Conference on Computer
  Communications}, 2021.

\bibitem{schreiber2013improved}
E.~L. Schreiber and R.~E. Korf, ``{Improved Bin Completion for Optimal Bin
  Packing and Number Partitioning},'' in \emph{Twenty-Third International Joint
  Conference on Artificial Intelligence}, 2013.

\bibitem{miao2017physical}
H.~Miao and M.~Faerber, ``Physical downlink control channel for 5g new radio,''
  in \emph{2017 European conference on networks and communications
  (EuCNC)}.\hskip 1em plus 0.5em minus 0.4em\relax IEEE, 2017, pp. 1--5.

\bibitem{yazar2018flexibility}
A.~Yazar and H.~Arslan, ``{A Flexibility Metric and Optimization Methods for
  Mixed Numerologies in 5G and Beyond},'' \emph{IEEE Access}, vol.~6, pp.
  3755--3764, 2018.

\end{thebibliography}
\end{document}